\RequirePackage{fix-cm}
\documentclass{svjour3}                     
\smartqed  
\usepackage{graphicx}
\journalname{}

\usepackage{amsmath}  
\usepackage{amssymb}  
\usepackage{amsfonts} 
\usepackage{graphicx} 
\newcommand {\bfv}[1] {{\boldsymbol {#1}}}

\newcommand\SEC[1]{}

\begin{document}

\title{Numerical study on the axisymmetric state in spherical Couette flow under unstable thermal stratification}

\author{
  Taishi Inagaki \and
  Tomoaki Itano  \and
  Masako Sugihara-Seki 
}

\institute{
  T. Inagaki \at
  Department of Pure and Applied Physics,\\
  Faculty of Engineering Science, Kansai University,  Osaka, 564-8680, Japan\\
  \and
  T. Itano \at
  Department of Pure and Applied Physics,\\
  Faculty of Engineering Science, Kansai University,  Osaka, 564-8680, Japan\\
  Tel.: +81-6-6368-1396\\
  \email{itano@kansai-u.ac.jp}           
  \and
  M. Sugihara-Seki \at         
  Department of Pure and Applied Physics,\\
  Faculty of Engineering Science, Kansai University,  Osaka, 564-8680, Japan\\
}

\date{Received: date / Accepted: date / Today: \today}

\maketitle





\begin{abstract}
{{This paper numerically investigates the shear flow between double concentric spherical boundaries rotating differentially, so-called {\it spherical Couette flow}, under unstable thermal stratification, focusing on the boundary of the axisymmetric/non-axisymmetric transition in wide gap cases where the inner radius is comparable to the clearance width.
While the transition of SCF has been confirmed experimentally in cases without thermal factor, insufficient knowledge on SCF subject to thermal instability, related to geophysical problems especially in wide gap cases, has been accumulated mainly based on numerical analysis; our motivation is to bridge the knowledge gap by a parameter extension.
We reconfirm that the transition under no thermal effect is initiated by a disturbance visualised as a spiral pattern with $n$ arms extending from the equatorial zone to the pole in each hemisphere, at the critical Reynolds number, ${\rm Re}_{\rm cr}$, as previously reported.
With increasing thermal factor, the buoyancy effect assists the system rotation to trigger a transition towards non-axisymmetric states, resulting in a relative decrease of ${\rm Re}_{\rm cr}$.
This is in contrast with the result that the system rotation apparently suppresses via Coriolis effect the transition to the thermally convective states at low Reynolds numbers.
The present study elucidates that the existence of the axisymmetric state is restricted within a closed area in the extended parameter space, along the boundary of which the spiral patterns observed experimentally in SCF continually connect to the classical spherical B\'enard convective states.
}}
\end{abstract}



\maketitle

\begin{flushleft}
{\large \bf Acknowledgement}  
\end{flushleft}
We thank Prof. Kageyama, Dr. Hori, Dr. Yokoyama, Prof. J. Seki, and Prof. N. Sugimoto for critical comments at earlier stages of the study.
T.I. expresses cordial thanks to Prof. Adachi for offering his numerical code based on the spectral element method to check the validity of our code.
T.I. is grateful for partial financial support from the Kansai University Subsidy for Supporting Young Scholars, 2016-2017, and an ORDIST group fund, 2018.
M.S. is grateful for partial financial support from KAKENHI 17H03176.
Finally, we would like to thank Editage for English language editing.

\section{Introduction}
\SEC{Polygonal Coherence - SCF (two control parameters, two cells)}
Rotating shear flow
 favours
 polygonal coherence around the poles.
Recent spacecraft missions have spotted the aspect of a regular polygon in the polar jet stream on Jupiter and Saturn\cite{God88}.
At the laboratory scale, liquid flow switches between axisymmetric and non-axisymmetric patterns in an open cylindrical vessel with either a stationary or rotating endwall and bottom \cite{Jan06,Tas09}. 
Here, we revisit yet another classical example, namely {\it spherical Couette flow} (SCF)---the incompressible Newtonian fluid confined between concentric spherical boundaries rotating differentially. SCF is a canonical yet practical turbulent shear flow.
As an idealised limit in the equatorial zone of the SCF, we can conceive the so-called planar Couette flow under periodic boundary conditions\cite{Ita09}.
The particular case between a rotating inner sphere and a stationary outer sphere, which has so far been investigated relatively extensively, is principally characterised by only two control parameters: {{a geometrical parameter and the Reynolds number, traditionally defined as ${\rm Re}=\frac{r_{\rm in}^2\Omega_{\rm in}}{\nu}$.}}
The basic laminar flow realised at a small ${\rm Re}$ consists of two zonal momentum cells in the north and south hemispheres, which are divided by a strong radial outward flow developing at the equatorial zone via the inertial (centrifugal) force of the rotating inner sphere.
A fluid element in one of the cells travels from the equator to a pole along the outer spherical boundary and then back to the equator on the rotating inner boundary, describing a spiral trajectory in the corresponding hemisphere.

\SEC{Distinction of transitions depending on $\beta$}
With an increase in the rotation rate of the inner sphere, SCF exhibits the first transition to turbulence.
Egbers and Rath \cite{Egb95} carried out experimental studies employing several spherical boundaries with {{different radii.
The geometry is determined by the ratio of the clearance width to the radius of the inner sphere $\beta=(r_{\rm out}-r_{\rm in})/r_{\rm in}$,}} where $r_{\rm in}$ and $r_{\rm out}$ are the inner and outer radii, respectively.
They noticed that the route to turbulence is determined by the value of $\beta$.
In ``{\it narrow gap}'' cases, where $\beta \le 0.25$, the transition is initiated by the axisymmetric Taylor vortices with successive instability, as observed in the cylindrical Taylor--Couette flow.
On the other hand, in ``{\it wide gap}'' cases, where $0.33 \le \beta$, the transition begins with a break of the polygonal secondary waves at a relatively high rotation rate, under the absence of Taylor vortices, as in the flow between two rotating planar disks.
The cases with $\beta=0.33$ and $\beta=0.5$ correspond to the onset of instability, which were visualised both as a sinuous wave disturbance at the equator propagating in the zonal direction, and as a spiral pattern with $n$ arms extending from the equatorial zone in each hemisphere.
Wulf {\it et al.} \cite{Wul99} further investigated successive transition processes undergoing several mode changes, and concluded that the Ruelle裕akens--Newhouse scenario, associated with a few polygonal modes, represents the route to turbulence.
The critical Reynolds numbers obtained were ${\rm Re}_{\rm cr}=2628$ ($n=6$) at $\beta=0.33$ and ${\rm Re}_{\rm cr}=1244$ ($n=5$) at $\beta=0.5$, the latter of which agrees remarkably well with a theoretical prediction based on linear stability analysis \cite{Ara97}.

\SEC{Shear vs. Heat}
Experimental studies on wide gap SCF promoted by a few pioneers during the last quarter of the past century\cite{Mun75,Bel91,Egb95} were originally stimulated by astronomical and geophysical problems.
The evidence of a fluid core in the planetary interior convinced researchers that thermal convection in the core is a prime mover inside an apparently solid yet pulsating planet, whereas direct measurements inside convection remain beyond our reach.
A framework to model the convection is Boussinesq fluid\cite{Cha61} in a spherical shell, which is the same as that employed in research on SCF.
By introducing electro-magnetic factors to the system \cite{Kid97,Sak99}, more realistic frameworks could be developed to elucidate the mechanism of secular variations in the terrestrial geomagnetic field.
Although these numerical models have been improved through comparisons with seismological or other tomographic data\cite{Fow04}, they are principally unrealisable in experiments, because it is impossible to reproduce the radial gravitational field except under a micro-gravity environment\cite{Feu11}.

\SEC{Homotopy of transition between SCF and SBC}
The motivation of the current work is to bring to bear the authenticity established on SCF---confirmed both numerically and experimentally---to our uncertain knowledge of {\it spherical B\'enard convection} (SBC) under the radial gravitational field, which has been accumulated only numerically.
For this purpose, we develop an SCF system with thermal and gravitational factors.
We first re-examine the transition in the SCF model at the vanishing limit of thermal factors.
By comparing our results to those of previous experiments on transition in the wide gap SCF, we validate the developed code.
We next examine the transition of the basic laminar flow perturbed by the {{thermal}} factor.
A phase diagram is obtained in the extended control parameter space that implies a cooperative and competitive mechanism between the thermal and inertia factors of this hybrid system.
{{
The remainder of this paper is organised as follows.
In the next section, we briefly describe the nondimensionalised governing equation of our system. 
In section 3, we reproduce the non-axisymmetric states obtained experimentally and discuss the morphology of vortex under shear and thermal effects.
In the latter part of the section, the diagram of equilibrium states of SCF is extended to the case under unstable thermal stratification.
The paper is concluded with some brief remarks on heat and angular momentum flux in the transition and the relevance of the present research.
}}

\begin{figure}[h]
  \centering
  \includegraphics[angle=0,width=0.53\columnwidth]{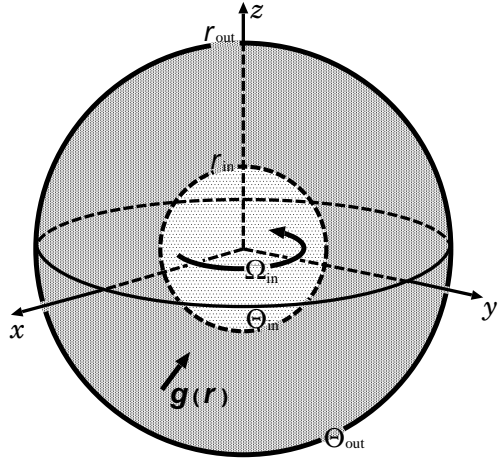}
  \label{config}
  \caption{
    Configuration of the present study. 
    The temperatures of the inner and outer spheres with radii $r_{\rm in}$ and $r_{\rm out}$ are fixed at $\Theta_{\rm in}$ and $\Theta_{\rm in}$, respectively.
    The inner sphere rotates at a constant angular velocity, $\Omega_{\rm in}$. 
    The primary parameters for SCF without the thermal effect are the geometrical parameter $\beta=(r_{\rm out}-r_{\rm in})/r_{\rm in}$, and the Reynolds number $\displaystyle{\rm Re}=\frac{r_{\rm in}^2\Omega_{\rm in}}{\nu}$. 
  }
\end{figure}


\section{Formulation}
We consider Boussinesq fluid confined between double concentric spherical boundaries, and we assume that the mass and internal heat source are distributed homogeneously inside the outer spherical boundary.
The dimensional equations for momentum and energy are
\[
\frac{{\rm D}\bfv{u}}{{\rm D}t} = - \frac{1}{\rho_0}\bfv{\nabla} p + \nu \bfv{\nabla}^2 \bfv{u} - \frac{\rho}{\rho_0} g_0 \bfv{r} \ \ , \ \ \frac{{\rm D}\Theta}{{\rm D}t} = \kappa \bfv{\nabla}^2 \Theta + \epsilon \ ,
\]
 where $\rho_0$, $\nu$, $g_0$, $\kappa$, and $\epsilon$ are the reference density, kinematic viscosity, ratio of gravitational acceleration to the radius, thermal diffusivity, and heat generation rate per specific heat\cite{Sak99}.
The temperature difference between the inner and outer spheres, principally determined by the heat generation rate $\epsilon$ and the radii, affects the momentum equation via the buoyancy term.
Here, half of the temperature difference between the inner and outer spherical boundaries, $\Delta \Theta=(\Theta_{\rm in}-\Theta_{\rm our})/2$, is determined by $\epsilon$.
The equations for the length, time, and temperature in the half gap width $\Delta r=(r_{\rm out}-r_{\rm in})/2$, the diffusion time $\Delta r^2/\nu$, and the temperature in $\Delta \Theta$, are nondimensionalised as follows:
\[
\frac{{\rm D}\bfv{u}}{{\rm D}t}=-\bfv{\nabla} p + \bfv{\nabla}^2 \bfv{u}+{\rm Gr} \Theta r \bfv{e}_r \ \ , \ \ 
\frac{{\rm D}\Theta}{{\rm D}t} = \frac{1}{{\rm Pr}} \Bigl( \bfv{\nabla}^2 \Theta + \frac{3\beta}{2+\beta} \Bigr) \ .
\]
Non-slip and isothermal boundary conditions are imposed at the stationary outer and rotating inner spherical boundaries.
The system is controlled by four parameters: $\displaystyle{\rm Re}=\frac{r_{\rm in}^2\Omega_{\rm in}}{\nu}$, $\displaystyle {\rm Pr}=\frac{\nu}{\kappa}$, and $\displaystyle {\rm Gr}=\frac{\alpha g_0 \Delta \Theta \Delta r^4}{\nu^2}$, as well as $\beta$.
At the vanishing limit of ${\rm Gr}$, the momentum equation is decoupled from the energy equation, where the system is equivalent to conventional SCF.

\SEC{Numerical procedure} 
We employ the Galerkin-spectral method to solve the governing equations numerically.
Because of the divergence-free constraint on $\bfv{u}$, we invoke toroidal and poloidal decomposition with regard to the radial direction, $\bfv{u}=\bfv{\nabla}\times ( \bfv{\nabla}\times (\Phi\bfv{r}) - \Psi \bfv{r})$.
Spatially expanding the scalar fields, $\Phi$, $\Psi$, and $\Theta$, in terms of Chebyshev polynomials and spherical harmonics with the aid of open libraries\cite{Fri05,Sch13}, and adapting the second-order Adams--Bashforth method complemented by Crank--Nikolson for temporal discretisation, the governing equation is reduced to an inhomogeneous Helmholtz equation.
After carefully confirming that the numerical solution is fully resolved (where the typical spatial resolution is $33\times 120\times 240$ modes in $r, \theta, \phi$ directions), we calculated the time evolution with time step $2^{-14}$.
Most equilibrium states reported hereafter, unless otherwise noted, were obtained by numerical integration initially beginning with a Stokes solution\cite{Lan87} with a small disturbance artificially generated by a series of random numbers.

\section{Results}
\begin{figure}[h]
  \begin{center}
    \includegraphics[width=0.89\columnwidth]{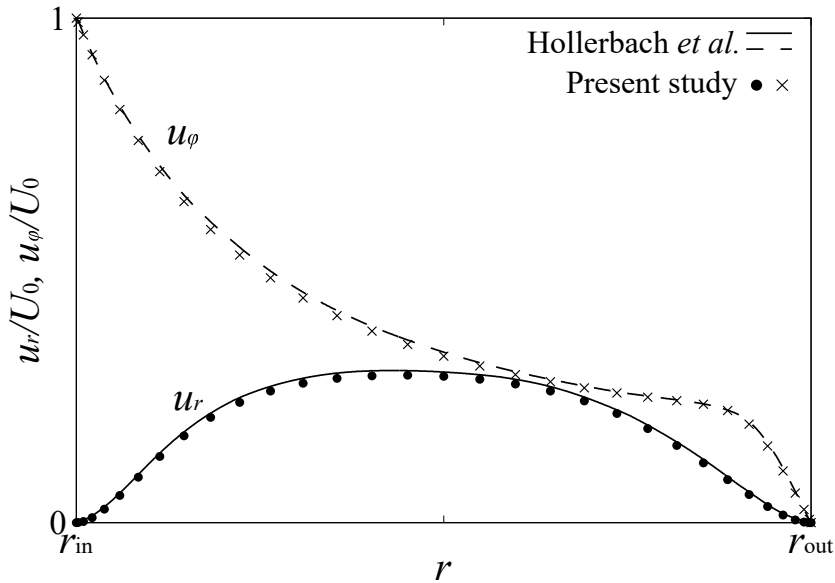}
  \end{center}
  \caption{ 
    Comparison of the basic laminar flow of SCF, $u_r$ ($\bfv{\bullet}$) and $u_\phi$ ($\bfv{\times}$) obtained at $(\beta,{\rm Re})=(1,489)$ (solid curve), to that of a previous numerical study (dashed curves), the value of which was read from the graph in {\em Fig. 1} of Ref.\cite{Hol06}.
  }
  \label{basic flow profile}
\end{figure}

\SEC{Validity with regard to basic flow}
The present numerical code was validated for ${\rm Re}<600$ in the case of SCF without the thermal factor, ${\rm Gr}=0$, by a comparison with previous experimental and numerical results\cite{Hol06,Nak02,Wul99}.
Fig.\ref{basic flow profile} shows the radial and zonal components of the basic laminar flow on the equatorial plane obtained at $(\beta,{\rm Re})=(1,489)$.
In the figure, the magnitudes of $u_r$ and $u_\phi$ are normalised by the zonal speed of the inner sphere at the equator, $U_0=\beta{\rm Re}/2$, which is equivalent to $r_{\rm in}\Omega_{\rm in}$ in dimensional variables.
Both the components agree with the previous numerical study\cite{Hol06} within an error of at most a few percentage points.
Note that the magnitude of the radial component is comparable to that of the zonal component at the centre of the boundaries.
As Hollerbach {\it et al.} \cite{Hol06} reported, this is characteristic in wide gap SCF but not in narrow gap cases.
In addition, due to the mixing effect of meridional circulation, the gradient of the zonal component localises in the layers on the boundaries as ${\rm Re}$ increases.
On the equatorial plane, strong radial outward flow forms the boundary layer of zonal flow near the outer sphere.

\begin{figure}[h]
  \begin{center}
    \includegraphics[width=0.94\columnwidth]{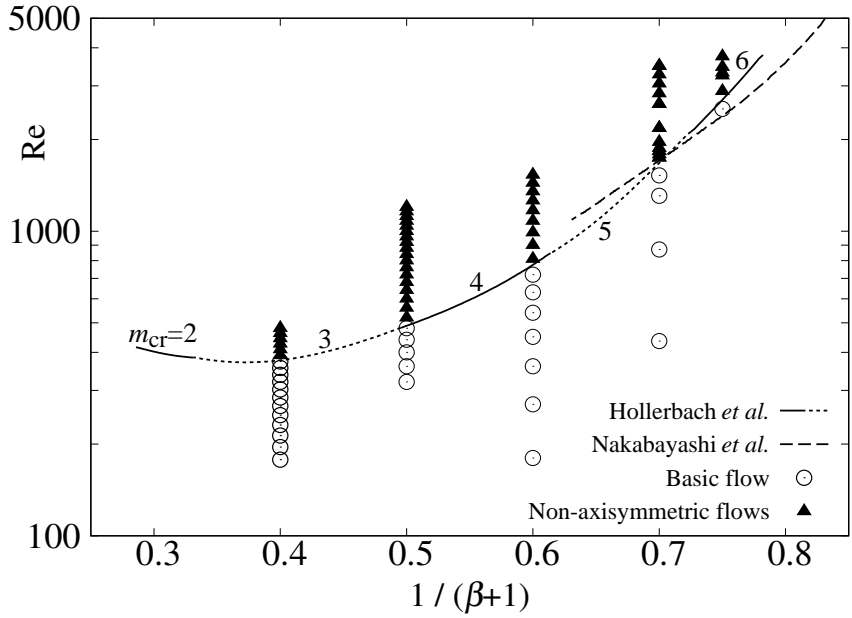}
  \end{center}
  \caption{ 
    Diagram of converged states with increasing Reynolds numbers for several ratios of the inner to outer radii $\eta$ ($\eta=1/(\beta+1)$).
    Narrow gap cases correspond to $0.75 \lesssim 1/(\beta+1) < 1$.
    Either the basic flow ($\circ$) or a non-axisymmetric state ($\blacktriangle$) is numerically obtained as a converged state at a set of $(\beta,{\rm Re})$. 
    The neutral curve against the cross-flow instability is indicated as the sequential solid/dotted curves at which non-axisymmetric equilibrium states emerge, as reported previously\cite{Hol06}. 
    The dashed curve is the cross-flow instability compiled in Ref.\cite{Nak02}.
  }
  \label{SCF transition}
\end{figure}

\SEC{Validity with regard to transition}
SCF undergoes the first transition from basic laminar flow to a non-axisymmetric state at a critical Reynolds number ${\rm Re}_{\rm cr}(\beta)$.
Using the developed code, we calculated a converged state starting from a (non-axisymmetric) small disturbance superimposed on the (axisymmetric) Stokes flow for a given $({\rm Re},\eta)$.
While a disturbance is attenuated at parameters (denoted by circles in Fig.\ref{SCF transition}), it is amplified and saturates (marked by triangles).
Additionally, the values of ${\rm Re}_{\rm cr}(\beta)$ obtained experimentally\cite{Wul99,Nak02} and numerically\cite{Hol06} are incorporated into the figure together with the present results.
The agreement between the present and the previous results is fairly good, except for a slight difference in the narrow gap case.

\SEC{Bound of wide gap}
We now discuss the lower bound of $\beta$ (the upper bound of $\eta$), 
{
 at a $\beta$ that is less the threshold at which the route to turbulence becomes distinct from that of wide gap SCF.
}
According to Nakabayashi {\it et al.}\cite{Nak02}, the definition of wide gap differs considerably depending on the researcher, ranging from $\beta_{\rm cr}=0.24$ ($\eta_{\rm cr}=0.81$)\cite{Mar87} to $\beta_{\rm cr}=0.48$ ($\eta_{\rm cr}=0.68$)\cite{Sch86}.
The divergence of ${\rm Re}(\beta)$ at $\beta\to \infty$ ($\eta \to 1$) indicates that the inner sphere diameter is inappropriate as a representative length scale to define the Reynolds number in narrow gap SCF, where instability originates at the shear stress between boundaries, as in planar Couette flow.
As pointed out in Ref.\cite{Hol06}, an alternative Reynolds number, defined as ${\rm Re}'=\beta {\rm Re}$, is more suitable for narrower cases of $\beta<\beta_{\rm cr}$ ($\eta>\eta_{\rm cr}$) than ${\rm Re}$.
In the present study, we focus on three wide-gap cases: $\beta=3/2, 1$, and $2/3$.

\SEC{Wave number}
A non-axisymmetric equilibrium state emerges at ${\rm Re}>{\rm Re}_{\rm cr}(\beta)$.
As the original system is homogeneous to the zonal direction, a wave number $m_{\rm cr}(\beta)$ should be selected at the onset of the instability, where an integer $m_{\rm cr}$ corresponds to the fundamental order in the spherical harmonic expansion activated due to the instability.
The perturbation of the equilibrium state consists of the fundamental mode and its higher harmonics, and { spontaneously} satisfies the following symmetry:
\begin{eqnarray*}
(\Phi,\Psi,\Theta)(r,\theta,\phi+\frac{2\pi}{m_{\rm cr}})&=&(\Phi,\Psi,\Theta)(r,\theta,\phi) \ \ ,\\  
 (\Phi,\Psi,\Theta)(r,\pi-\theta,\phi)&=&(-\Phi,\Psi,-\Theta)(r,\theta,\phi) \ \ .
\end{eqnarray*}
The wave number $m_{\rm cr}(\beta)$ is $3$, $4$, and $5$ for $\beta=3/2$, $1$, and $2/3$, respectively, as indicated beside the individual solid curves in Fig.\ref{SCF transition}.

\SEC{Sinuous wave at the equator}
The second symmetry is reminiscent of the so-called shift-reflection symmetry predominant in the planar Couette flow for low Reynolds numbers\cite{Ham95}.
The low-speed streaky structure prevalent in the near-wall region of wall-bounded shear flows, first discovered experimentally by Kline {\it et al.}\cite{Kli67}, is generated by wall-normal lift flow sustained by streamwise vortices\cite{Wal97}.
However, note that the wall-normal component is not inherent in the basic parallel flows in such turbulent shear flows on the plane wall.
The shift-reflection symmetry of the disturbance is favourable under shear stress between sliding parallel walls, leading to a meandering of the low-speed streak, visualised experimentally in the near-wall region.
In SCF, the bifurcated equilibrium state is visualised as a sinuous wave pattern at the equator propagating to the zonal direction (not shown here).

\begin{figure}[h]
  \begin{center}
    \includegraphics[width=0.65\columnwidth]{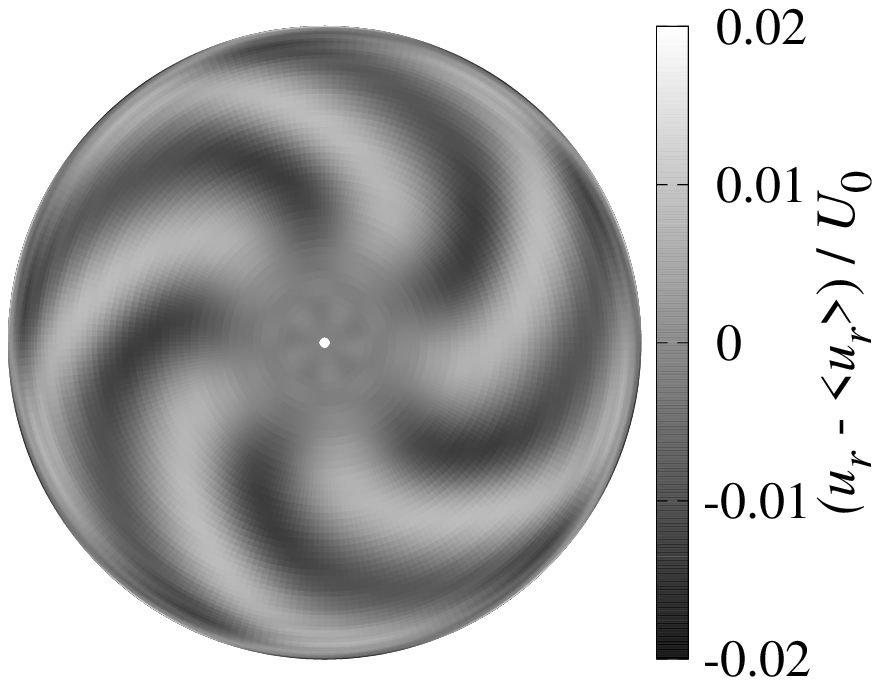}
  \end{center}
  \caption{ 
    Contour of radial component normalised by $r_{\rm in}\Omega_{\rm in}$ for the polar hemisphere at $(\eta,{\rm Re},{\rm Gr})=(2/3,1320,0)$:
  }
  \label{pentagon visualisation}
\end{figure}

\begin{figure}[h]
  \begin{center}
    \includegraphics[width=0.30\columnwidth]{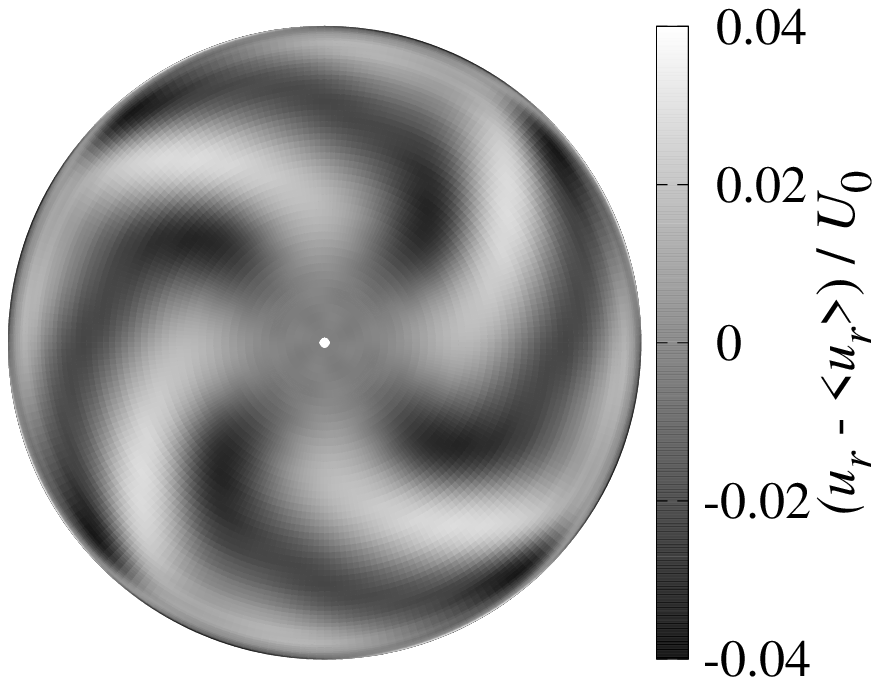}
    \includegraphics[width=0.30\columnwidth]{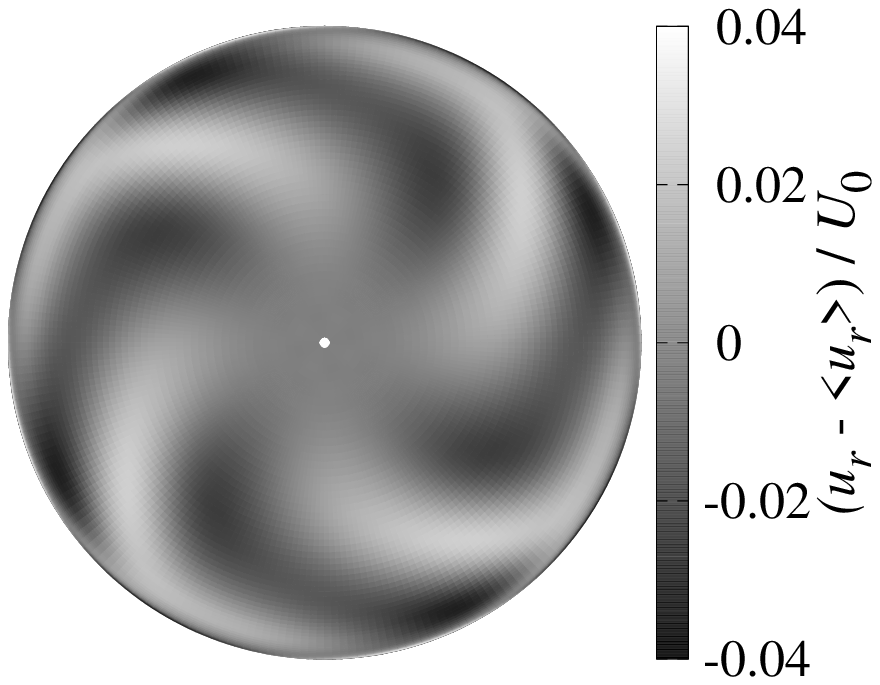}
    \includegraphics[width=0.30\columnwidth]{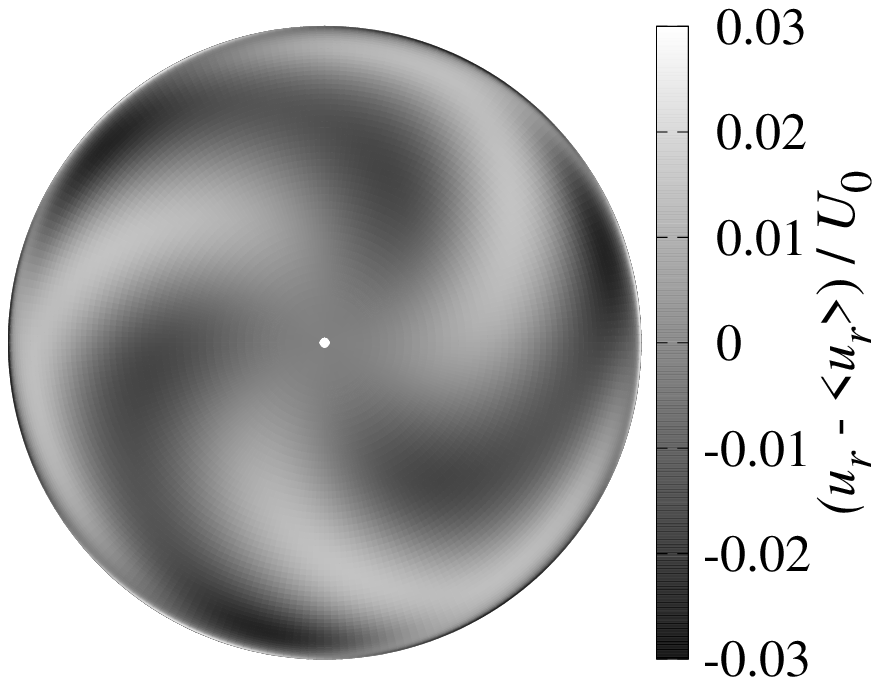}
  \end{center}
  \caption{ 
    Contour of radial component normalised by $r_{\rm in}\Omega_{\rm in}$ for the polar hemisphere at $(\beta,{\rm Re})=(2/3,900)$ (left), $(1,520)$ (centre), and $(3/2,391)$ (right).
    Polar hemisphere at $(\beta,{\rm Re})=(1,1320)$
  }
  \label{spirals}
\end{figure}

\SEC{Spiral in a hemisphere}
As stipulated by the first symmetry, the non-axisymmetric equilibrium state that emerges over ${\rm Re}_{\rm cr}$ has a spiral pattern with $m_{\rm cr}$ { vortical} arms alternately extending from the equator to the poles in each hemisphere{, which is superimposed on the basic flow}.
Fig.\ref{pentagon visualisation} provides a qualitative aspect in the polar hemisphere of the spiral pattern $(\eta,{\rm Re})=(2/3,1320)$ of the present result, which shows an agreement to an experimental visualisation by Wulf {\it et al.} (see Fig. 9(a) in Ref.\cite{Wul99}).
Here, we define the operator $\langle f \rangle = \frac{1}{2\pi r\sin{(\theta)}} \int_0^{2\pi} f(r,\theta,\phi) d\phi$. The present result is visualised as a contour of the radial component of perturbed velocity, $u_r-\langle u_r \rangle$ at $r=r_0${, while the spiral pattern in the experimental visualisation in Ref.\cite{Wul99} is illuminated with a suspention of Iriodin 111 platelets.}
At the onset of the instability of basic laminar flow, we numerically confirm that the number of arms is equivalent to $m_{\rm cr}(\beta)$, as shown in Fig.\ref{spirals}.

\SEC{Change of the number of arms}
Further numerical investigations for higher Reynolds numbers indicated that the number of arms either increases or decreases, with an increase of ${\rm Re}$.
Focusing on the initial stage of the transition, the number of arms decreases as ${\rm Re}$ increases.
The number of arms in the spiral pattern varies from $4$ to $3$ at {\em ${\rm Re}=560$} for $\beta=1$ ($\eta=0.5$).
Here, we emphasise again that the initial condition used is the Stokes solution with a small disturbance.
In a range of ${\rm Re}$ 
 before the transition toward a spatiotemporally irregular state, we confirmed that different equilibrium states are realised from different initial states at the same Reynolds number.
This is associated with the non-uniqueness of the equilibrium states realisable via hysteresis over ${\rm Re}_{\rm cr}$, as reported in Ref.\cite{Egb95}, which have been investigated in detail numerically \cite{Abb18a}.
Hereafter, we refer to the non-axisymmetric equilibrium state with $m$ spiral arms from a pole as an ``$m$-fold spiral state''.

\begin{figure}[h]
  \begin{center}
\includegraphics[width=0.70\columnwidth]{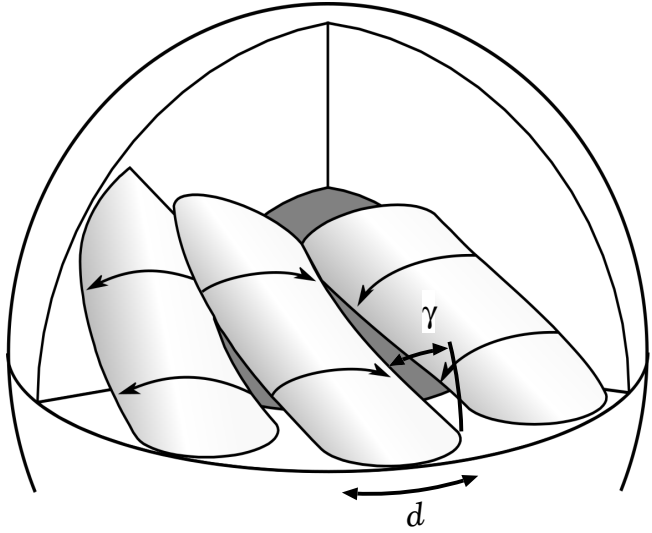}
    \end{center}
    \caption{ 
      Schematic view of a non-axisymmetric disturbance in an $m$-fold spiral state. 
      The axis of the structure is inclined by $\gamma$ against the meridional plane.
      In a strict sense, the adjacent vortical structures extending to the north and south poles cancel out each other on the equatorial plane, because of the symmetry satisfied by $\Phi$.
      }
    \label{a vortical structure}
\end{figure}
\SEC{Speculation on the inclination}
In general, the instantaneous velocity field $\bfv{u}$ may be decomposed into the axisymmetric component $\bfv{u}_{m=0}(r,\theta)$ and non-axisymmetric component $\bfv{u}_{m\ne 0}(r,\theta,\phi)$.
The non-axisymmetric component, which is enhanced due to the instability of the axisymmetric component beyond ${\rm Re}_{\rm cr}$, describes a disturbance of $m$ pairs of positive and negative vortical structures extending from each pole.
Assuming that each vortical structure is circular along the cross-section vertical to the vortex axis, and that it spans the entire gap between the spherical boundaries $2 \Delta r$, we can estimate the system circumference at the equatorial section as the product of the number of vortical structures and an effective arc-length of a vortical structure projected onto the equatorial plane $d$ ($d> 2\Delta r$)---i.e. $2 \pi r_0 \approx 2 m \times d$.
If the axis of a vortical structure is inclined by $\gamma$ with respect to the meridional plane, then $\frac{2 \Delta r}{d} \approx \cos{\gamma}$ (see Fig.\ref{a vortical structure}).
We suppose that the inclination is estimated from the advection of $\bfv{u}_{m=0}(r,\theta)$, such that the incline angle $\gamma$ is estimated by $\tan{\gamma} \approx \frac{U_\phi}{U_\theta}$, where $U_{r}$, $U_\theta$ and $U_\phi$ are the representative magnitudes of $\bfv{u}_{m=0}(r_0,\theta)$ maximised in the equatorial zone.
Additionally, $U_{r}$ is comparable to $U_{\theta}$ on the equatorial plane.
The above speculation is followed by
\[
 m \approx \frac{\pi }{2} (\frac{2}{\beta}+1) \Bigl( \sqrt{1+ \frac{U_{0}^2}{U_{r}^2} } \ \Bigr)^{-1} \ \ .
\]
In principle, $m$ increases as $\beta$ decreases (the gap becomes narrower).
At the limit of the narrow gap, the magnitude of $U_{r}/U_{0}$ vanishes, as reported in Ref.\cite{Hol06}, and it is probably on the higher order of $1/\beta$.
Thus, the axisymmetric Taylor vortex flow ($\gamma=\pi/2$) is the trigger for the onset of the transition and $m\to 0$.
On the other hand, in wide gap SCF, $U_r$ is principally on the order of $U_{0}$, as seen in Fig.\ref{basic flow profile}, such that the value of $m$ is the inverse of the geometrical factor $\beta$.
Although the increase of ${\rm Re}$ enhances the meridional circulation of $\bfv{u}_{m=0}$, the increase of $U_{r}$ is weakened due to mixing in the gap induced by the non-axisymmetric component, once ${\rm Re}$ passes beyond ${\rm Re}_{\rm cr}$.
Thus, beyond ${\rm Re}_{\rm cr}$, a further increase of ${\rm Re}$ reduces the ratio of $U_{r}$ to $U_{0}$, which leads to a decrease of $m$.

\begin{figure}
  \begin{center}
   \includegraphics[width=0.90\columnwidth]{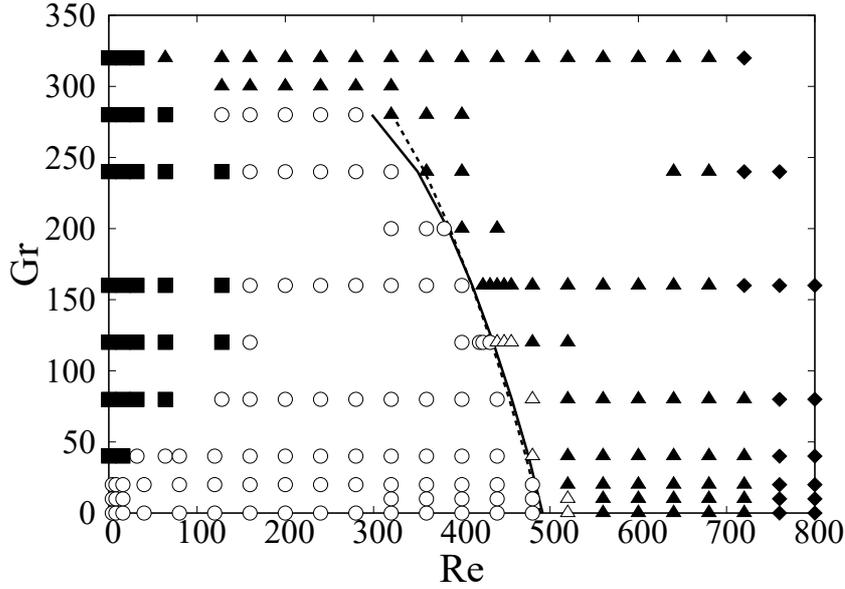}
    \end{center}
    \caption{ 
      Equilibrium states in ${\rm Re}$-${\rm Gr}$ space:  basic state ($\circ$), four-fold spiral state ($\triangle$), three-fold spiral state ($\blacktriangle$), aperiodic state ($\blacklozenge$), and heat convective state ($\blacksquare$).
      }
    \label{eta0.5}
\end{figure}
\SEC{Increase of ${\rm Gr}$}
By introducing the thermal factor into the SCF by increasing ${\rm Gr}$, we extend the phase diagram of SCF up to ${\rm Gr}\le 300$ for the case where $\beta=1$.
In the present study, our scope is restricted in the case of ${\rm Pr}=1$.
Equilibrium states obtained from the present numerical integration are classified here into a {\em basic state, three-fold spiral state, four-fold spiral state, periodic states, and heat convective states}, indicated by the symbols shown in Fig.\ref{eta0.5}.
Because the aforementioned non-uniqueness is inevitable in the system, the obtained equilibrium state may be preferred, yet it is not the only state but rather one realisable state with a given set of parameters.
While faster computation is desirable when investigating a broad range of this parameter space, more time is needed to ensure that the numerically converged state is the equilibrium.
We ensure convergence by monitoring several state variables that reflect the axisymmetry and the mirror-symmetry on the equatorial plane.
The average interval time required for convergence is of the order of ten in a nondimensional time-scale.

\SEC{Review on ${\rm Re=0}$}
In Fig.\ref{eta0.5}, the cases where ${\rm Re}=0$ correspond to the classical Rayleigh--B\'enard problem in spherical geometry.
Unstable modes that have degenerated owing to spherical homogeneity evolve simultaneously from the static thermal conductive state ($\bfv{u}=0$) at the critical {\em ${\rm Gr}_{\rm cr}=37.36$} with a zenith degree $l_{\rm cr}=4$ at ${\rm Re}=0$, as shown in Ref.\cite{Ita15}.
From nonlinear interactions between some eigenmodes, highly symmetric steady states invariant under a set of transformations of point groups---e.g. axisymmetric or polyhedral patterns---can bifurcate directly from the static state.
Several symmetric patterns that emerge over the critical Grashof number have been explored since the previous pioneering research\cite{Cha61,Bus75,Zeb83}. However, a novel less-symmetric rotating state still attracts attention\cite{Li10,Sig11,Ita15}.
Therefore, due to the non-uniqueness in the range above the critical Grashof number, the final equilibrium state depends on the initial condition.

\SEC{In small ${\rm Re}$ range}
A slight increase in ${\rm Re}$ breaks the isotropic symmetry of the achieved states ${\rm Re}=0$.
Thus, the meridional flow induced by the rotating inner sphere converts the static thermal conductive state to an axisymmetric basic state, at a { fixed} ${\rm Gr}$ that is less than the critical Grashof number.
Note that the axisymmetric basic state exists as an equilibrium state in the entire parameter space, even if it is unstable over the critical Grashof number.
With the introduction of ${\rm Re}$, the eigenmodes degenerate at ${\rm Gr}>{\rm Gr}_{\rm cr}$ and ${\rm Re}=0$ separates.
However, instability originating at eigenmodes with $m\ne 0$ is suppressed by the meridional flow induced by the rotating inner sphere.
Thus, an increase in ${\rm Re}$ leads to an apparent increase in the critical Grashof number ${\rm Gr}_{\rm cr}(\beta,{\rm Re})$, over which the nonaxisymmetricity typical in heat convective states triggers the instability.
The most dangerous eigenmode at a critical Grashof number with a small ${\rm Re} \approx 0$ value is the axisymmetric mode with $l=4$, which, in the meridional cross-section, generates the radial currents intensified locally at the equator and at the poles.
The dependence of the critical Grashof number on ${\rm Re}$ could be accounted for by the Coriolis factor.
The dependency associated with the Coriolis factor was { previously} calculated for a rotating spherical shell (in the case of inner and outer spheres rotating with the same angular velocity) with a homogeneously distributed heat source\cite{Cha61}{, where} introducing this Coriolis term into the momentum equation { led} to a qualitative estimation: ${\rm Gr}_{\rm cr} -{\rm Gr}_{\rm cr}(0) \propto {\rm Re}^2$.
Such a relation in the present SCF-SBC system might provide the explicit boundary between heat convective states and the basic laminar flow in Fig.\ref{eta0.5}.

\SEC{In large ${\rm Re}$ range}
Thus, around a critical ${\rm Gr}$ at ${\rm Re=0}$, an increase of ${\rm Re}$ apparently stabilises the system.
On the contrary, in general, shear stress and unstable thermal stratification intensify with an increase of ${\rm Re}$ and ${\rm Gr}$, respectively, and this should contribute to the instability of the system.
One might consider that with, an increase of ${\rm Gr}$, the buoyancy force enforces radial outward flow induced by the inertial force of the rotating inner sphere at the equator.
Such speculation leads to an estimation with regard to the boundary of the basic laminar flow in the control parameter space: i.e. $k {\rm Gr}_{\rm cr} + \Omega_{\rm in}^2 \approx {\rm Const}$.
In Fig.\ref{eta0.5}, this relationship is reflected qualitatively as a convex boundary curve originating at ${\rm Re}=489$, when the effective ratio of the buoyancy to the inertial forces, $k \approx 8$, is adopted as a fitting parameter.
Thus, in the range of a relatively large ${\rm Re}$, the increase of ${\rm Re}$ leads to a decrease of the critical Grashof number ${\rm Gr}_{\rm cr}(\beta,{\rm Re})$, over which the most dominant state is $n$-fold states typical in normal SCF.
This was confirmed for other wide-gap cases: viz. $\beta=3/2$ and $2/3$.

\section{Discussion}
\SEC{Number of arms at ${\rm Gr}({\rm Re})$}
It is of interest that as ${\rm Re}$ decreases, the obtained equilibrium state over ${\rm Gr}_{\rm cr}(\beta,{\rm Re})$ changes from a three-fold (in SCF, ${\rm Re}\sim 489$) to a four-fold spiral state (in the SCF-SBC system ${\rm Re} \lesssim 400$).
Additionally, by linearising the nonlinear terms with respect to the non-axisymmetric modes, we calculated the linear growth rates of a (non-axisymmetric) small disturbance superimposed on the (axisymmetric) basic laminar flow for a given $({\rm Re},\eta)$.
The boundary of positive/negative growth $m=3$ and $m=4$ is plotted with dashed and solid curves in Fig.\ref{eta0.5}, respectively.
The exchange of the dominant fold state on the neutral curve ${\rm Gr}={\rm Gr}_{\rm cr}(\beta,{\rm Re})$ occurs at the intersection of these curves {, which is likely to be in $120<{\rm Gr}<160$ and $400<{\rm Re}<440$ in the figure}.

The exchange in the number of arms on the neutral curve could be explained qualitatively from the aforementioned speculation.
If the increase in ${\rm Gr}$ contributes to enforcing the radial outward flow, then, the increase of $U_{\theta}$ would result in a larger $m$.
However, in reality, at a ${\rm Gr}$ beyond ${\rm Gr}_{\rm cr}({\rm Re})$ under a given ${\rm Re}$, the further increase of ${\rm Gr}$ probably activates the zenith degree of freedom in a polar zone where the radial flow is relatively weak.
The system with $\beta=1$ principally prefers a wave number of four as the critical meridional wave number in thermal convection, so the selection of an $m=3$ regime could survive in a delicate balance of competitive features between a shear effect at the equator and a thermal effect around the poles.

\begin{figure}[h]
  \begin{center}
    \includegraphics[width=0.80\columnwidth]{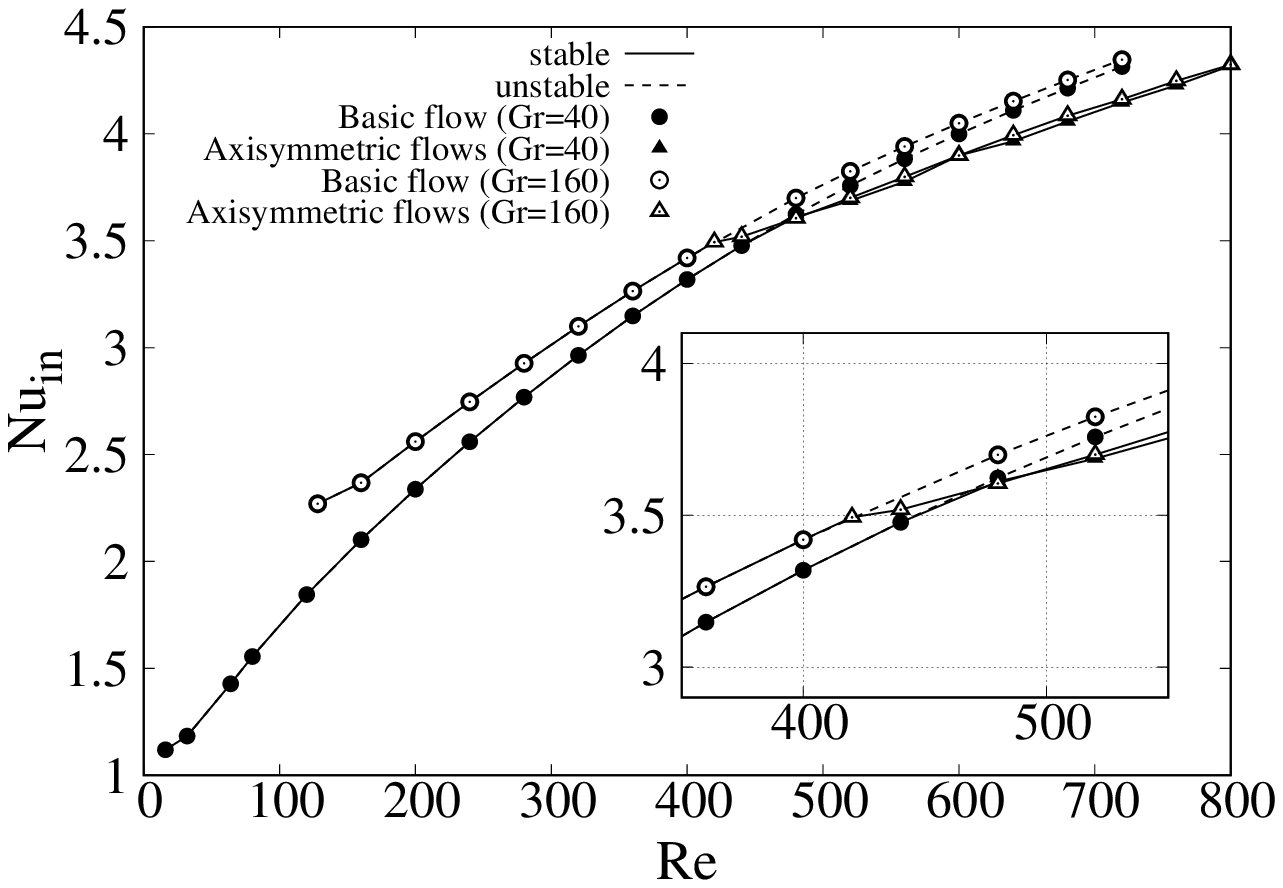}
    \includegraphics[width=0.80\columnwidth]{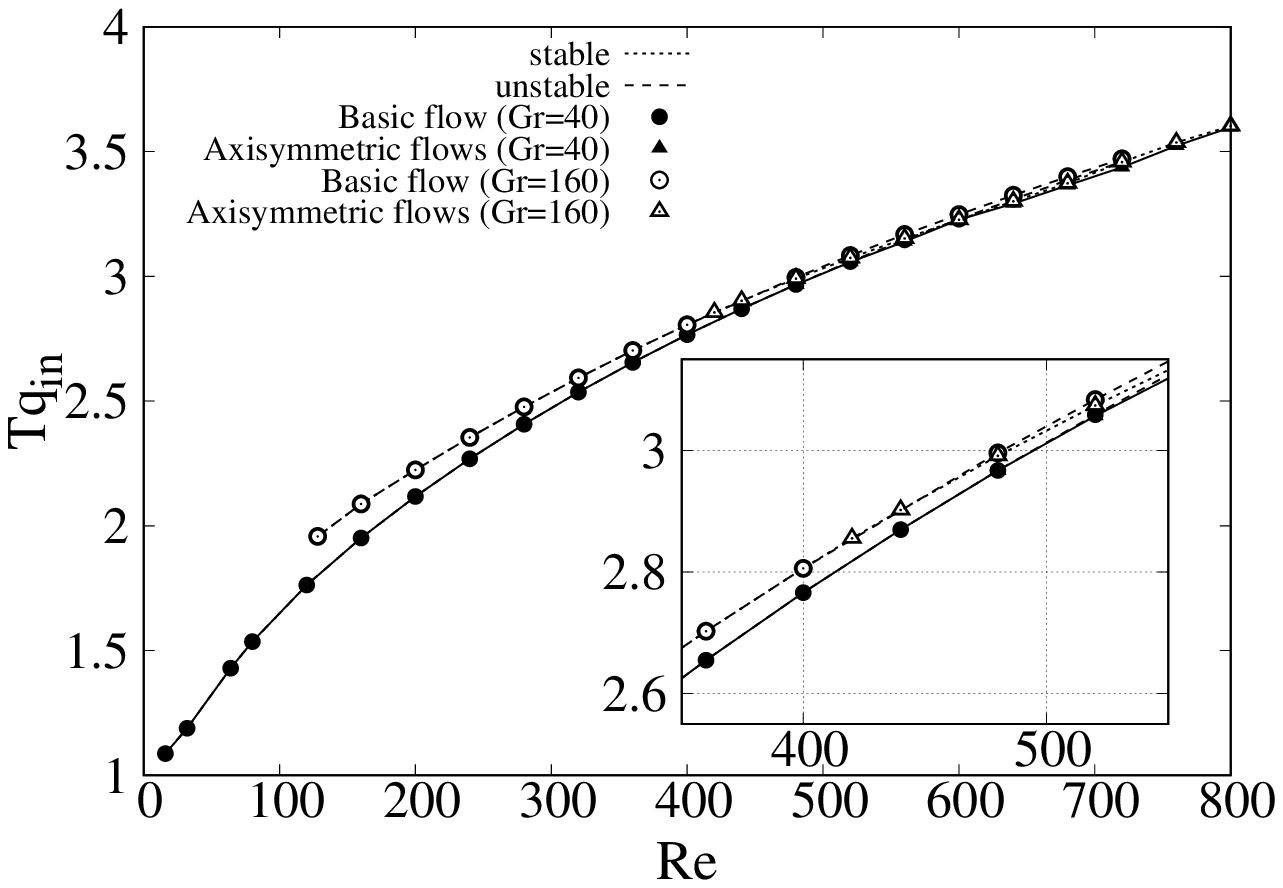}
    \end{center}
    \caption{ 
      Heat and angular momentum fluxes transferred radially outward from the inner sphere, for equilibrium states obtained at ${\rm Gr}=40$ (filled symbols) and $160$ (unfilled symbols):
      (top) heat energy ${\rm Nu}_{\rm in}$, and (bottom) angular momentum ${\rm Tq}_{\rm in}$.
      The $m$-fold states (represented by filled and unfilled triangles) bifurcate from the basic flow (represented by filled and unfilled circles) over a critical Reynolds number.
      The magnitudes of the fluxes of the $m$-fold states are less than those of the basic flow.
      }
    \label{Nu_Tq}
\end{figure}

\SEC{Torque \& Nusselt}
The present system transfers heat energy radially outwards from the inner sphere, which can be enhanced further by the fluid convection.
The heat flux normalised by that of the Stokes solution is conventionally termed the Nusselt number, ${\rm Nu}_{\rm in}$, calculated by
\[ {\rm Nu}_{\rm in}= 1-\frac{\rm Pr}{4\pi r_{\rm in}} \int\frac{\partial \Theta}{\partial r}  \Bigr|_{r=r_{\rm in}} \sin{\theta} d\theta d\phi \ \ .\]
At the same time, the rotating inner sphere also transfers angular momentum outwards, which can likewise be enhanced by the fluid motion.
The magnitude of angular momentum flux can be evaluated by the $z$ component of the torque exerted on the inner sphere normalised by that of the Stokes solution, ${\rm Tq}_{\rm in}$, and calculated exclusively from $\Psi$.
Figs.\ref{Nu_Tq}(a) and (b) show the ${\rm Nu}_{\rm in}$ and ${\rm Tq}_{\rm in}$ of the equilibrium states obtained at ${\rm Gr}=40$ and $160$, respectively.
With an increase of ${\rm Re}$, these magnitudes increase monotonically, but the increases are attenuated for ${\rm Re}$, satisfying ${\rm Gr}>{\rm Gr}_{\rm cr}({\rm Re})$.
From the perspective of determinism, the obtained equilibrium state should be selected exclusively from the initial condition and the values of control parameters.
On the other hand, the possible selection of a converged state in a meta-stable system depends on how large the basin attraction of the state is in the phase space.
Thus, one would expect that the system prefers a state with { either larger heat transfer or more angular-momentum transfer\cite{Kit06}.}
However, the present results do not support this argument, because both the ${\rm Nu}_{\rm in}$ and ${\rm Tq}_{\rm in}$ of the non-axisymmetric state are smaller than those of the basic laminar state for ${\rm Gr} > {\rm Gr}({\rm Re})$.


\SEC{Akinaga \& Araki}
A phase diagram in the ${\rm Re}-{\rm Gr}$ parameter space has been investigated numerically for another thermal convection induced by instabilities under the presence of transverse seepage \cite{Aki16}.
Air flow was modelled in a double-paned window with a vertically inflowing bottom as the planar Poiseuille flow between vertical parallel plates kept at different temperatures.
The control parameters of the system were ${\rm Re}$ and ${\rm Gr}$, to evaluate the ratios of inertia and of buoyancy to viscosity.
Steady equilibrium states bifurcated from the basic flow were obtained numerically using the Newton--Raphson method.
The authors showed that the bifurcation of the secondary travelling wave solution can be either supercritical or subcritical in the parameter space spanned by ${\rm Re}$ and ${\rm Gr}$.
Whether our system is a subcritical system remains to be determined in future work.

\section{Conclusion}
The spherical Couette flow under unstable thermal stratification was numerically investigated for $\eta=0.5$.
In the case without thermal stratification (${\rm Gr}=0$), the axisymmetric/non-axisymmetric transition was initiated by $n$-fold spiral states at a critical Reynolds number, the results for which agreed with those obtained experimentally\cite{Wul99,Nak02} and numerically\cite{Hol06}.
Next, we investigated the dependency on the Reynolds number, ${\rm Re}$, of the critical Grashof number ${\rm Gr}_{\rm cr}$ at which the axisymmetric state loses its stability against non-axisymmetric disturbance.
At a small Reynolds number, the rotation of the inner sphere suppresses the transition, where (axisymmetric) basic laminar flow bifurcates into a variety of states under the thermal effect, such that ${\rm Gr}_{\rm cr}$ increases with ${\rm Re}$.
On the other hand, at larger Reynolds numbers, the rotation enhances the transition due to the effect of inertia, where the basic laminar state bifurcates into an $n$-fold spiral state, such that ${\rm Gr}_{\rm cr}$ decreases with ${\rm Re}$.
{
The axisymmetric state exists in a closed parameter range of the extended parameter space, along the boundary of which the spiral patterns observed experimentally in SCF continually connect to the classical spherical B\'enard convective states.
}
This implies the existence of an upper limit of ${\rm Gr}_{\rm cr}$ at an optimum value of ${\rm Re}$.

{
  The present study can extend the diagram of equilibrium states of SCF, while the non-uniqueness of solutions still remains an open question.
  Particularly, the non-uniqueness in a range above the critical Reynolds number under ${\rm Gr}=0$, which was confirmed by experiments, will be solved more rigorously only by numerical bifurcation analysis.
  The analysis would also distinguish the static and the thermally bifurcated solutions above the critical Grashof number under ${\rm Re}=0$, which are both axisymmetric in a strict sense.
  This will explicitly provide the boundary between heat convective states and the basic laminar flow in Fig.\ref{eta0.5}, the details for which will be investigated in our future work.

}

{
  In this study, Nusselt number and torque on the inner sphere are also calculated on the obtained axisymmetric and non-axisymmetric states over the critical Grashof number.
  In the extended system, the state which is realised most probably among possible steady states corresponds to neither the one with maximum heat transfer nor the one with maximum angular-momentum transfer hypothesized by the principle of maximum entropy.
  The natural explanation for the enhancement of the Nusselt number validated in simple thermal convection is not applicable to the extended system.
}

\bibliographystyle{apsrev4-1}
\bibliography{scf02}

\begin{thebibliography}{10}%
\makeatletter
\providecommand \@ifxundefined [1]{%
 \ifx #1\undefined \expandafter \@firstoftwo
 \else \expandafter \@secondoftwo
\fi
}%
\providecommand \@ifnum [1]{%
 \ifnum #1\expandafter \@firstoftwo
 \else \expandafter \@secondoftwo
\fi
}%
\providecommand \enquote [1]{``#1''}%
\providecommand \bibnamefont  [1]{#1}%
\providecommand \bibfnamefont [1]{#1}%
\providecommand \citenamefont [1]{#1}%
\providecommand\href[0]{\@sanitize\@href}%
\providecommand\@href[1]{\endgroup\@@startlink{#1}\endgroup\@@href}%
\providecommand\@@href[1]{#1\@@endlink}%
\providecommand \@sanitize [0]{\begingroup\catcode`\&12\catcode`\#12\relax}%
\@ifxundefined \pdfoutput {\@firstoftwo}{%
 \@ifnum{\z@=\pdfoutput}{\@firstoftwo}{\@secondoftwo}%
}{%
 \providecommand\@@startlink[1]{\leavevmode\special{html:<a href="#1">}}%
 \providecommand\@@endlink[0]{\special{html:</a>}}%
}{%
 \providecommand\@@startlink[1]{%
  \leavevmode
  \pdfstartlink
   attr{/Border[0 0 1 ]/H/I/C[0 1 1]}%
   user{/Subtype/Link/A<</Type/Action/S/URI/URI(#1)>>}%
  \relax
 }%
 \providecommand\@@endlink[0]{\pdfendlink}%
}%
\providecommand \url  [0]{\begingroup\@sanitize \@url }%
\providecommand \@url [1]{\endgroup\@href {#1}{\urlprefix}}%
\providecommand \urlprefix [0]{URL }%
\providecommand \Eprint[0]{\href }%
\@ifxundefined \urlstyle {%
  \providecommand \doi [1]{doi:\discretionary{}{}{}#1}%
}{%
  \providecommand \doi [0]{doi:\discretionary{}{}{}\begingroup
  \urlstyle{rm}\Url }%
}%
\providecommand \doibase [0]{http://dx.doi.org/}%
\providecommand \Doi[1]{\href{\doibase#1}}%
\providecommand \bibAnnote [3]{%
  \BibitemShut{#1}%
  \begin{quotation}\noindent
    \textsc{Key:}\ #2\\\textsc{Annotation:}\ #3%
  \end{quotation}%
}%
\providecommand \bibAnnoteFile [2]{%
  \IfFileExists{#2}{\bibAnnote {#1} {#2} {\input{#2}}}{}%
}%
\providecommand \typeout [0]{\immediate \write \m@ne }%
\providecommand \selectlanguage [0]{\@gobble}%
\providecommand \bibinfo [0]{\@secondoftwo}%
\providecommand \bibfield [0]{\@secondoftwo}%
\providecommand \translation [1]{[#1]}%
\providecommand \BibitemOpen[0]{}%
\providecommand \bibitemStop [0]{}%
\providecommand \bibitemNoStop [0]{.\EOS\space}%
\providecommand \EOS [0]{\spacefactor3000\relax}%
\providecommand \BibitemShut [1]{\csname bibitem#1\endcsname}%
\bibitem{God88}%
  \BibitemOpen
  \bibfield{author}{%
  \bibinfo {author} {\bibnamefont{D.A.Godfrey}},\ }%
  \bibfield{journal}{%
  \bibinfo {journal} {Icarus}\ }%
  \textbf{\bibinfo {volume} {76}},\ \bibinfo {pages} {335} (\bibinfo {year}
  {1988})%
  \bibAnnoteFile{NoStop}{God88}%
\bibitem{Jan06}%
  \BibitemOpen
  \bibfield{author}{%
  \bibinfo {author} {\bibfnamefont{T.~R.~N.}\ \bibnamefont{Jansson}}, \bibinfo
  {author} {\bibfnamefont{M.~P.}\ \bibnamefont{Haspang}}, \bibinfo {author}
  {\bibfnamefont{K.~H.}\ \bibnamefont{Jensen}}, \bibinfo {author}
  {\bibfnamefont{P.}~\bibnamefont{Hersen}},\ and\ \bibinfo {author}
  {\bibfnamefont{T.}~\bibnamefont{Bohr}},\ }%
  \bibfield{journal}{%
  \bibinfo {journal} {Phys. Rev. Lett.}\ }%
  \textbf{\bibinfo {volume} {96}},\ \bibinfo {pages} {174502} (\bibinfo {year}
  {2006})%
  \bibAnnoteFile{NoStop}{Jan06}%
\bibitem{Tas09}%
  \BibitemOpen
  \bibfield{author}{%
  \bibinfo {author} {\bibfnamefont{Y.}~\bibnamefont{Tasaka}}\ and\ \bibinfo
  {author} {\bibfnamefont{M.}~\bibnamefont{Iima}},\ }%
  \bibfield{journal}{%
  \bibinfo {journal} {J. Fluid Mech.}\ }%
  \textbf{\bibinfo {volume} {636}},\ \bibinfo {pages} {475} (\bibinfo {year}
  {2009})%
  \bibAnnoteFile{NoStop}{Tas09}%
\bibitem{Ita09}%
  \BibitemOpen
  \bibfield{author}{%
  \bibinfo {author} {\bibfnamefont{T.}~\bibnamefont{Itano}}\ and\ \bibinfo
  {author} {\bibfnamefont{S.~C.}\ \bibnamefont{Generalis}},\ }%
  \bibfield{journal}{%
  \bibinfo {journal} {Phys. Rev. Lett.}\ }%
  \textbf{\bibinfo {volume} {102}},\ \bibinfo {pages} {114501} (\bibinfo {year}
  {2009})%
  \bibAnnoteFile{NoStop}{Ita09}%
\bibitem{Egb95}%
  \BibitemOpen
  \bibfield{author}{%
  \bibinfo {author} {\bibfnamefont{C.}~\bibnamefont{Egbers}}\ and\ \bibinfo
  {author} {\bibfnamefont{H.~J.}\ \bibnamefont{Rath}},\ }%
  \bibfield{journal}{%
  \bibinfo {journal} {Acta Mechanica}\ }%
  \textbf{\bibinfo {volume} {111}},\ \bibinfo {pages} {125} (\bibinfo {year}
  {1995})%
  \bibAnnoteFile{NoStop}{Egb95}%
\bibitem{Wul99}%
  \BibitemOpen
  \bibfield{author}{%
  \bibinfo {author} {\bibfnamefont{P.}~\bibnamefont{Wulf}}, \bibinfo {author}
  {\bibfnamefont{C.}~\bibnamefont{Egbers}},\ and\ \bibinfo {author}
  {\bibfnamefont{H.~J.}\ \bibnamefont{Rath}},\ }%
  \bibfield{journal}{%
  \bibinfo {journal} {Phys. Fluids}\ }%
  \textbf{\bibinfo {volume} {11}},\ \bibinfo {pages} {1359} (\bibinfo {year}
  {1999})%
  \bibAnnoteFile{NoStop}{Wul99}%
\bibitem{Ara97}%
  \BibitemOpen
  \bibfield{author}{%
  \bibinfo {author} {\bibfnamefont{K.}~\bibnamefont{Araki}}, \bibinfo {author}
  {\bibfnamefont{J.}~\bibnamefont{Mizushima}},\ and\ \bibinfo {author}
  {\bibfnamefont{S.}~\bibnamefont{Yanase}},\ }%
  \bibfield{journal}{%
  \bibinfo {journal} {Phys. Fluids}\ }%
  \textbf{\bibinfo {volume} {9}},\ \bibinfo {pages} {1197} (\bibinfo {year}
  {1997})%
  \bibAnnoteFile{NoStop}{Ara97}%
\bibitem{Mun75}%
  \BibitemOpen
  \bibfield{author}{%
  \bibinfo {author} {\bibfnamefont{B.~R.}\ \bibnamefont{Munson}}\ and\ \bibinfo
  {author} {\bibfnamefont{M.}~\bibnamefont{Menguturk}},\ }%
  \bibfield{journal}{%
  \bibinfo {journal} {J. Fluid Mech.}\ }%
  \textbf{\bibinfo {volume} {69}},\ \bibinfo {pages} {705} (\bibinfo {year}
  {1975})%
  \bibAnnoteFile{NoStop}{Mun75}%
\bibitem{Bel91}%
  \BibitemOpen
  \bibfield{author}{%
  \bibinfo {author} {\bibfnamefont{Y.~N.}\ \bibnamefont{Belyaef}}\ and\
  \bibinfo {author} {\bibfnamefont{I.~M.}\ \bibnamefont{Yavorskaya}},\ }%
  \bibfield{journal}{%
  \bibinfo {journal} {(translated from) Izvestiya Akademii Nauk SSSR, Mekh.
  Zhid. i Gaza}\ }%
  \textbf{\bibinfo {volume} {1}},\ \bibinfo {pages} {10} (\bibinfo {year}
  {1991})%
  \bibAnnoteFile{NoStop}{Bel91}%
\bibitem{Cha61}%
  \BibitemOpen
  \bibfield{author}{%
  \bibinfo {author} {\bibfnamefont{S.}~\bibnamefont{Chandrasekhar}},\ }%
  \emph{\bibinfo {title} {Hydrodynamic and Hydromagnetic Stability}}\ (\bibinfo
  {publisher} {Clerendon Press, Oxford},\ \bibinfo {year} {1961})%
  \bibAnnoteFile{NoStop}{Cha61}%
\bibitem{Kid97}%
  \BibitemOpen
  \bibfield{author}{%
  \bibinfo {author} {\bibfnamefont{S.}~\bibnamefont{Kida}}, \bibinfo {author}
  {\bibfnamefont{K.}~\bibnamefont{Araki}},\ and\ \bibinfo {author}
  {\bibfnamefont{H.}~\bibnamefont{Kitauchi}},\ }%
  \bibfield{journal}{%
  \bibinfo {journal} {J. Phys. Soc. Jpn}\ }%
  \textbf{\bibinfo {volume} {66(7)}},\ \bibinfo {pages} {2194} (\bibinfo {year}
  {1997})%
  \bibAnnoteFile{NoStop}{Kid97}%
\bibitem{Sak99}%
  \BibitemOpen
  \bibfield{author}{%
  \bibinfo {author} {\bibfnamefont{A.}~\bibnamefont{Sakuraba}}\ and\ \bibinfo
  {author} {\bibfnamefont{M.}~\bibnamefont{Kono}},\ }%
  \bibfield{journal}{%
  \bibinfo {journal} {Physics of the Earth and Planetary Interiors}\ }%
  \textbf{\bibinfo {volume} {1111}},\ \bibinfo {pages} {105} (\bibinfo {year}
  {1999})%
  \bibAnnoteFile{NoStop}{Sak99}%
\bibitem{Fow04}%
  \BibitemOpen
  \bibfield{author}{%
  \bibinfo {author} {\bibfnamefont{C.~M.~R.}\ \bibnamefont{Fowler}},\ }%
  \emph{\bibinfo {title} {The Solid Earth}}\ (\bibinfo {publisher} {Cambridge
  University Press},\ \bibinfo {year} {2004})%
  \bibAnnoteFile{NoStop}{Fow04}%
\bibitem{Feu11}%
  \BibitemOpen
  \bibfield{author}{%
  \bibinfo {author} {\bibfnamefont{F.}~\bibnamefont{Feudel}}, \bibinfo {author}
  {\bibfnamefont{K.}~\bibnamefont{Bergemann}}, \bibinfo {author}
  {\bibfnamefont{L.}~\bibnamefont{Tuckerman}}, \bibinfo {author}
  {\bibfnamefont{C.}~\bibnamefont{Egbers}}, \bibinfo {author}
  {\bibfnamefont{B.}~\bibnamefont{Futterer}}, \bibinfo {author}
  {\bibfnamefont{M.}~\bibnamefont{Gellert}},\ and\ \bibinfo {author}
  {\bibfnamefont{R.}~\bibnamefont{Hollerbach}},\ }%
  \bibfield{journal}{%
  \bibinfo {journal} {Phys. Rev. E}\ }%
  \textbf{\bibinfo {volume} {83}},\ \bibinfo {pages} {046304} (\bibinfo {year}
  {2011})%
  \bibAnnoteFile{NoStop}{Feu11}%
\bibitem{Fri05}%
  \BibitemOpen
  \bibfield{author}{%
  \bibinfo {author} {\bibfnamefont{M.}~\bibnamefont{Frigo}}\ and\ \bibinfo
  {author} {\bibfnamefont{S.~G.}\ \bibnamefont{Johnson}},\ }%
  \bibfield{journal}{%
  \bibinfo {journal} {Proceedings of the IEEE}\ }%
  \textbf{\bibinfo {volume} {93}},\ \bibinfo {pages} {216} (\bibinfo {year}
  {2005}),\ \bibinfo {note} {special issue on ``Program Generation,
  Optimization, and Platf orm Adaptation''}%
  \bibAnnoteFile{NoStop}{Fri05}%
\bibitem{Sch13}%
  \BibitemOpen
  \bibfield{author}{%
  \bibinfo {author} {\bibfnamefont{N.}~\bibnamefont{Schaeffer}},\ }%
  \bibfield{journal}{%
  \bibinfo {journal} {Geochemistry, Geophysics, Geosystems}\ }%
  \textbf{\bibinfo {volume} {14}},\ \bibinfo {pages} {751} (\bibinfo {year}
  {2013})%
  \bibAnnoteFile{NoStop}{Sch13}%
\bibitem{Lan87}%
  \BibitemOpen
  \bibfield{author}{%
  \bibinfo {author} {\bibfnamefont{L.~D.}\ \bibnamefont{Landau}}\ and\ \bibinfo
  {author} {\bibfnamefont{E.}~\bibnamefont{Lifshitz}},\ }%
  \emph{\bibinfo {title} {Fluid Mechanics: Vol.6 (Course of Theoretical
  Physics) 2nd Edition}}\ (\bibinfo {publisher} {Butterworth-Heinemann},\
  \bibinfo {year} {1987})%
  \bibAnnoteFile{NoStop}{Lan87}%
\bibitem{Hol06}%
  \BibitemOpen
  \bibfield{author}{%
  \bibinfo {author} {\bibfnamefont{R.}~\bibnamefont{Hollerbach}}, \bibinfo
  {author} {\bibfnamefont{M.}~\bibnamefont{Junk}},\ and\ \bibinfo {author}
  {\bibfnamefont{C.}~\bibnamefont{Egbers}},\ }%
  \bibfield{journal}{%
  \bibinfo {journal} {Fluid Dyanmics Research}\ }%
  \textbf{\bibinfo {volume} {38}},\ \bibinfo {pages} {257} (\bibinfo {year}
  {2006})%
  \bibAnnoteFile{NoStop}{Hol06}%
\bibitem{Nak02}%
  \BibitemOpen
  \bibfield{author}{%
  \bibinfo {author} {\bibfnamefont{K.}~\bibnamefont{Nakabayashi}}, \bibinfo
  {author} {\bibfnamefont{Y.}~\bibnamefont{Tsuchida}},\ and\ \bibinfo {author}
  {\bibfnamefont{Z.}~\bibnamefont{Zheng}},\ }%
  \bibfield{journal}{%
  \bibinfo {journal} {Phys. Fluids}\ }%
  \textbf{\bibinfo {volume} {14}},\ \bibinfo {pages} {3963} (\bibinfo {year}
  {2002})%
  \bibAnnoteFile{NoStop}{Nak02}%
\bibitem{Mar87}%
  \BibitemOpen
  \bibfield{author}{%
  \bibinfo {author} {\bibfnamefont{P.~S.}\ \bibnamefont{Marcus}}\ and\ \bibinfo
  {author} {\bibfnamefont{L.~S.}\ \bibnamefont{Tuckerman}},\ }%
  \bibfield{journal}{%
  \bibinfo {journal} {J. Fluid Mech.}\ }%
  \textbf{\bibinfo {volume} {185}},\ \bibinfo {pages} {1} (\bibinfo {year}
  {1987})%
  \bibAnnoteFile{NoStop}{Mar87}%
\bibitem{Sch86}%
  \BibitemOpen
  \bibfield{author}{%
  \bibinfo {author} {\bibfnamefont{G.}~\bibnamefont{Schrauf}},\ }%
  \bibfield{journal}{%
  \bibinfo {journal} {J. Fluid Mech.}\ }%
  \textbf{\bibinfo {volume} {166}},\ \bibinfo {pages} {287} (\bibinfo {year}
  {1986})%
  \bibAnnoteFile{NoStop}{Sch86}%
\bibitem{Ham95}%
  \BibitemOpen
  \bibfield{author}{%
  \bibinfo {author} {\bibfnamefont{J.~M.}\ \bibnamefont{Hamilton}}, \bibinfo
  {author} {\bibfnamefont{J.}~\bibnamefont{Kim}},\ and\ \bibinfo {author}
  {\bibfnamefont{F.}~\bibnamefont{Waleffe}},\ }%
  \bibfield{journal}{%
  \bibinfo {journal} {Phys. Fluids}\ }%
  \textbf{\bibinfo {volume} {287}},\ \bibinfo {pages} {317} (\bibinfo {year}
  {1995})%
  \bibAnnoteFile{NoStop}{Ham95}%
\bibitem{Kli67}%
  \BibitemOpen
  \bibfield{author}{%
  \bibinfo {author} {\bibfnamefont{S.~J.}\ \bibnamefont{Klines}}, \bibinfo
  {author} {\bibfnamefont{W.}~\bibnamefont{Reynolds}}, \bibinfo {author}
  {\bibnamefont{F.A.Schraub}},\ and\ \bibinfo {author}
  {\bibnamefont{P.W.Runstadler}},\ }%
  \bibfield{journal}{%
  \bibinfo {journal} {J.~Fluid Mech.}\ }%
  \textbf{\bibinfo {volume} {30}},\ \bibinfo {pages} {741} (\bibinfo {year}
  {1967})%
  \bibAnnoteFile{NoStop}{Kli67}%
\bibitem{Wal97}%
  \BibitemOpen
  \bibfield{author}{%
  \bibinfo {author} {\bibfnamefont{F.}~\bibnamefont{Waleffe}},\ }%
  \bibfield{journal}{%
  \bibinfo {journal} {Phys. Fluids}\ }%
  \textbf{\bibinfo {volume} {9}},\ \bibinfo {pages} {883} (\bibinfo {year}
  {1997})%
  \bibAnnoteFile{NoStop}{Wal97}%
\bibitem{Abb18a}%
  \BibitemOpen
  \bibfield{author}{%
  \bibinfo {author} {\bibfnamefont{S.}~\bibnamefont{Abbas}}, \bibinfo {author}
  {\bibfnamefont{L.}~\bibnamefont{Yuan}},\ and\ \bibinfo {author}
  {\bibfnamefont{A.}~\bibnamefont{Shah}},\ }%
  \bibfield{journal}{%
  \bibinfo {journal} {Fluid Dyn. Res.}\ }%
  \textbf{\bibinfo {volume} {50}},\ \bibinfo {pages} {025507} (\bibinfo {year}
  {2018})%
  \bibAnnoteFile{NoStop}{Abb18a}%
\bibitem{Ita15}%
  \BibitemOpen
  \bibfield{author}{%
  \bibinfo {author} {\bibfnamefont{T.}~\bibnamefont{Itano}}, \bibinfo {author}
  {\bibfnamefont{T.}~\bibnamefont{Ninomiya}}, \bibinfo {author}
  {\bibfnamefont{K.}~\bibnamefont{Konno}},\ and\ \bibinfo {author}
  {\bibfnamefont{M.}~\bibnamefont{Sugihara-Seki}},\ }%
  \bibfield{journal}{%
  \bibinfo {journal} {J. Phys. Soc. Japan}\ }%
  \textbf{\bibinfo {volume} {84}},\ \bibinfo {pages} {103401} (\bibinfo {year}
  {2015})%
  \bibAnnoteFile{NoStop}{Ita15}%
\bibitem{Bus75}%
  \BibitemOpen
  \bibfield{author}{%
  \bibinfo {author} {\bibfnamefont{F.~H.}\ \bibnamefont{Busse}},\ }%
  \bibfield{journal}{%
  \bibinfo {journal} {J. Fluid Mech.}\ }%
  \textbf{\bibinfo {volume} {72(1)}},\ \bibinfo {pages} {67} (\bibinfo {year}
  {1975})%
  \bibAnnoteFile{NoStop}{Bus75}%
\bibitem{Zeb83}%
  \BibitemOpen
  \bibfield{author}{%
  \bibinfo {author} {\bibfnamefont{A.}~\bibnamefont{Zebib}}, \bibinfo {author}
  {\bibfnamefont{G.}~\bibnamefont{Schubert}}, \bibinfo {author}
  {\bibfnamefont{J.}~\bibnamefont{Dein}},\ and\ \bibinfo {author}
  {\bibfnamefont{R.}~\bibnamefont{Pariwal}},\ }%
  \bibfield{journal}{%
  \bibinfo {journal} {Geophys. Astrophys. Fluid Dyn.}\ }%
  \textbf{\bibinfo {volume} {23}},\ \bibinfo {pages} {1} (\bibinfo {year}
  {1983})%
  \bibAnnoteFile{NoStop}{Zeb83}%
\bibitem{Li10}%
  \BibitemOpen
  \bibfield{author}{%
  \bibinfo {author} {\bibfnamefont{L.}~\bibnamefont{Li}}, \bibinfo {author}
  {\bibfnamefont{X.}~\bibnamefont{Liao}}, \bibinfo {author}
  {\bibfnamefont{K.~H.}\ \bibnamefont{Chan}},\ and\ \bibinfo {author}
  {\bibfnamefont{K.}~\bibnamefont{Zhang}},\ }%
  \bibfield{journal}{%
  \bibinfo {journal} {Phys. Fluids}\ }%
  \textbf{\bibinfo {volume} {22}},\ \bibinfo {pages} {011701} (\bibinfo {year}
  {2010})%
  \bibAnnoteFile{NoStop}{Li10}%
\bibitem{Sig11}%
  \BibitemOpen
  \bibfield{author}{%
  \bibinfo {author} {\bibfnamefont{R.}~\bibnamefont{Sigrist}}\ and\ \bibinfo
  {author} {\bibfnamefont{P.}~\bibnamefont{Matthews}},\ }%
  \bibfield{journal}{%
  \bibinfo {journal} {SIAM J. Applied Dynamical Systems}\ }%
  \textbf{\bibinfo {volume} {10(3)}},\ \bibinfo {pages} {1177} (\bibinfo {year}
  {2011})%
  \bibAnnoteFile{NoStop}{Sig11}%
\bibitem{Kit06}%
  \BibitemOpen
  \bibfield{author}{%
  \bibinfo {author} {\bibfnamefont{T.}~\bibnamefont{Kita}},\ }%
  \bibfield{journal}{%
  \bibinfo {journal} {J. Phys. Soc. Japan}\ }%
  \textbf{\bibinfo {volume} {75}},\ \bibinfo {pages} {124005} (\bibinfo {year}
  {2006})%
  \bibAnnoteFile{NoStop}{Kit06}%
\bibitem{Aki16}%
  \BibitemOpen
  \bibfield{author}{%
  \bibinfo {author} {\bibfnamefont{T.}~\bibnamefont{Akinaga}}, \bibinfo
  {author} {\bibfnamefont{T.}~\bibnamefont{Itano}},\ and\ \bibinfo {author}
  {\bibfnamefont{S.}~\bibnamefont{Generalis}},\ }%
  \bibfield{journal}{%
  \bibinfo {journal} {Chaos, Solitons and Fractals}\ }%
  \textbf{\bibinfo {volume} {91}},\ \bibinfo {pages} {533} (\bibinfo {year}
  {2016})%
  \bibAnnoteFile{NoStop}{Aki16}%
\end{thebibliography}%

\end{document}